\documentstyle[emulateapj,psfig,apjfonts]{article}

\def\sig8{\mathrel{\sigma_{\rm 8}}}
\def\ls{\mathrel{\hbox{\rlap{\hbox{\lower4pt\hbox{$\sim$}}}\hbox{$<$}}}}
\def\gs{\mathrel{\hbox{\rlap{\hbox{\lower4pt\hbox{$\sim$}}}\hbox{$>$}}}}
\def\om0{\mathrel{\Omega_{\rm o}}}
\def\Lo{\mathrel{\Lambda_{\rm o}}}
\def\Ho{\mathrel{H_{\rm o}}}

\def\kms{\mathrel{\rm km\,s^{-1}}}
\def\Msol{\mathrel{\rm M_{\odot}}}

\def\hkpc{\mathrel{h^{-1}{\rm kpc}}}
\def\hMpc{\mathrel{h^{-1}{\rm Mpc}}}

\def\Mpc{\mathrel{\rm Mpc}}

\def\lx{\mathrel{L_{\rm X}}}
\def\tx{\mathrel{T_{\rm X}}}
\def\keV{\mathrel{{\rm keV}}}
\def\txtot{\mathrel{T_{\rm X,tot}}}
\def\txann{\mathrel{T_{\rm X,ann}}}
\def\Mtot{\mathrel{M_{\rm tot}}}
\def\Msub{\mathrel{M_{\rm sub}}}

\def\ergs{\mathrel{\rm erg\,s^{-1}}}
\def\tcool{\mathrel{t_{\rm cool}}}

\def\chandra{\emph{Chandra$\ $}}

\lefthead{Smith et al.}
\righthead{Impact of Cluster Substructure on {\Large $\sig8$}}

\begin{document}
\title{Measuring {\Large $\sig8$} with Cluster Lensing : Biases from
unrelaxed clusters}

\author{
Graham P.\ Smith,$\!$\altaffilmark{1,2}
Alastair C.\ Edge,$\!$\altaffilmark{3}
Vincent R.\ Eke,$\!$\altaffilmark{3}
Robert C.\ Nichol,$\!$\altaffilmark{4}
Ian Smail,$\!$\altaffilmark{3} \&
Jean-Paul Kneib,$\!$\altaffilmark{2,5}
}

\setcounter{footnote}{0}

\altaffiltext{1}{Department of Physics, University of Durham, South
  Road, Durham DH1 3LE, UK}
\altaffiltext{2}{Current address: California Institute of Technology,
  MC 105-24, Pasadena, CA 91125, USA -- Email: gps@astro.caltech.edu} 
\altaffiltext{3}{Institute for Computational Cosmology, Department of
  Physics, University of Durham, South Road, Durham DH1 3LE, UK}
\altaffiltext{4}{Department of Physics, Carnegie Mellon University,
5000 Forbes Avenue, Pittsburgh, PA 15213-3890, USA}
\altaffiltext{5}{Observatoire Midi-Pyr\'en\'ees, 14 Avenue E.\,Belin, 
31400 Toulouse, France}
\setcounter{footnote}{5}

\begin{abstract}
We use gravitational lens models and X-ray spectral analysis of ten
X-ray luminous galaxy clusters at $z\simeq0.2$ to study the impact of
cluster substructure on attempts to normalize the matter power
spectrum.  We estimate that unrelaxed clusters are 30\% hotter than
relaxed clusters causing $\sig8$ to be over-estimated by 20\% if the
cluster selection function is not accounted for correctly.  This helps
to explain the wide range in $\sig8$ derived from different
techniques, $\sig8\sim0.6$--1, and offers a physically motivated
explanation for some of the discrepancy.  We also identify two further
systematics in our analysis: (i) extrapolation of small field-of-view
mass measurements to the cluster virial radius and (ii) projection of
3--dimensional masses contained in numerical simulations to the
2--dimensional information that is available from observations.  We
combine quantitative estimates of these two effects with our model
fitting to estimate from the current data that $\sig8=0.75\pm0.05({\rm
statistical})\pm0.15({\rm systematic})$, where the systematic error
reflects the extrapolation and projection uncertainties.  All three
systematics (substructure, extrapolation and projection) are
fundamental to future cluster-based measurements of $\sig8$ regardless
of the techniques employed.  However, we identify gravitational
lensing as the tool of choice for such studies, because a combination
of strong- and weak-lensing offers the most direct route to control
the systematics and thus achieve an unbiased comparison between
observation and theory.
\end{abstract}


\section{Introduction}

The spectrum of cosmic matter fluctuations is an important constraint
on theoretical models of structure formation (e.g.\ Press \& Schechter
1974; Bond et al.\ 1991; Bower 1991; Kauffmann \& White 1993; Lacey \&
Cole 1993).  The amplitude of the power spectrum is parametrized as
$\sig8$, the linear-theory value of the rms fractional fluctuations in
density averaged in spheres of $8\hMpc$ radius at $z=0$.  Several
methods have been used to estimate $\sig8$: measurement of the
abundance of galaxy clusters (Eke, Cole \& Frenk 1996; Borgani et al.\
2001; Reiprich \& B\"ohringer 2001; Seljak 2001; Allen et al.\ 2002;
Viana, Nichol \& Liddle 2002; Pierpaoli et al.\ 2001), cosmic shear
analyses (e.g.\ Bacon et al.\ 2002; Hoekstra et al.\ 2002; Refregier
et al.\ 2002; van Waerbeke et al.\ 2002; Brown et al.\ 2002; Jarvis et
al.\ 2002), cosmic microwave background (CMB) studies (Sievers et al.\
2002; Bond et al.\ 2002), combined analysis of galaxy redshift survey
and CMB data (Lahav et al.\ 2002).  Current estimates of $\sig8$ range
from $\sim0.6$ to $\sim1.0$, with claimed statistical uncertainties in
the range $\Delta\!\sig8\,\sim0.02$--0.15.  Overall, the situation is
characterized by a lack of agreement between the results from
different methods, or the same methods used on different samples,
suggesting that systematic uncertainties probably lie at the heart of
the current disagreement over the value of $\sig8$.

In this letter we investigate systematic biases in the use of
cluster abundances to measure $\sig8$.  In principal the mass function
of galaxy clusters, $n(>\!M)$, should yield a direct constraint on
$\sig8$.  However, it is not currently possible to measure cluster
masses with the precision and in the numbers required to construct a
robust cluster mass function from direct measurement.  The local
cluster X-ray temperature function, $n(>\!T)$, has proved more
accessible (e.g.\ Edge et al.\ 1990; Henry \& Arnaud 1991; Markevitch
1998; Blanchard et al.\ 2000; Pierpaoli et al.\ 2000; Ikebe et al.\
2002).  The X-ray temperature function in conjunction with a robust
mass-temperature calibration therefore offers an opportunity to
constrain $\sig8$.

Observational attempts to calibrate the mass-temperature relation
typically rely on X-ray observations of clusters (e.g.\ Markevitch
1998; Nevalainen, Markevitch \& Forman 2000; Finoguenov, Reiprich \&
B\"ohringer 2001; Allen, Schmidt \& Fabian 2001, hereafter ASF; Reiprich \&
B\"ohringer 2002).  Despite the progress made by Allen (1998) in
understanding X-ray based cluster mass measurements, X-ray techniques
continue to assume that all clusters are symmetric, equilibrium
systems.  This is a major concern, because $\sim40$--70\% of galaxy
clusters appear to be dynamically immature (e.g.\ Mohr et al.\ 1995;
Buote \& Tsai 1996; Ota \& Mitsuda 2002; Smith et al.\ 2003, in prep.,
hereafter S03), and this immaturity has a measurable systematic impact on the
normalization of the cluster mass-temperature and mass-luminosity
relations (Ota \& Mitsuda 2002; S03; see also Randall et al.\ 2002).

In contrast, mass estimates based on gravitational lensing are
insensitive to the physical nature and state of the cluster mass.
Cluster lensing studies are therefore free from the symmetry and
equilibrium assumptions that plague the X-ray studies.  Attempts to
use lensing to calibrate the cluster mass-temperature relation have so
far relied on previously published and/or crude cluster mass estimates
(Hjorth et al.\ 1998; ASF; Viana, Nichol \& Liddle 2002).  A major
improvement on these pioneering studies would come from a precise and
uniform analysis of a large objectively selected cluster sample for
which high resolution space-based optical and X-ray data were
available.  In anticipation of such a program we conduct a pilot study
using S03's \emph{Hubble Space Telescope (HST)/Chandra} gravitational
lensing survey of ten X-ray luminous galaxy clusters at $z\simeq0.2$.
S03 made precise cluster mass and temperature measurements and thus
constrained the high-mass end of the cluster mass-temperature
relation.  They also studied the dependence of this normalization on
cluster substructure, concluding that unrelaxed clusters are, on
average, $30\%$ hotter than relaxed clusters.  S03's results therefore
offer a unique opportunity to study the impact of cluster substructure
on estimates of $\sig8$.

We summarize S03's results in \S2, describe our modeling and results
in \S3 and summarize our conclusions in \S4.  We express the Hubble
parameter in terms of $h$, where $\Ho=100\,h\kms\Mpc^{-1}$.  We also
adopt $\om0=0.3$ and
$\Lo=0.7$.

\section{\emph{HST/Chandra} Mass-Temperature Calibration}

S03 studied a representative sample of ten of the most X-ray luminous
clusters ($\lx\ge2\times 10^{44}h^{-2}\ergs$, 0.1--2.4\,keV) in a
narrow redshift slice at $z=0.21\pm0.04$, with line-of-sight reddening
of $E(B-V)\le0.1$ from the XBACs sample (Ebeling et al.\ 1996).  Each
cluster was typically observed for 3 orbits (i.e.\ 7.5ks) through the
F702W filter using the WFPC2 camera onboard \emph{HST}.  S03 used
these data in conjunction with ground-based optical and near-infrared
data (Smith et al.\ 2001, 2002), and the {\sc lenstool} software
(Kneib 1993; Kneib et al.\ 1996; Smith 2002) to construct a detailed
gravitational lens model of each cluster.

Armed with these models, S03 measured $M_{2500}$, the total projected
cluster mass within $r_{2500}$, i.e.\ the radius at which the density
of matter in the clusters falls to $\rho=\rho_{2500}=2500\rho_{\rm
c}$, where $\rho_{\rm c}$ is the critical density required to close
the universe.\footnote{At $z=0.2$, $r_{2500}$ corresponds to the edge
of the \emph{HST}/WFPC2 field of view for the most massive clusters in
S03's sample.}  S03 also used the models to divide their sample into
relaxed ($\Msub\!/\!\Mtot<10\%$) and unrelaxed
($\Msub\!/\!\Mtot>10\%$) clusters where $\Mtot$ is the total projected
mass of the cluster within $r_{\rm 2500}$ and $\Msub$ is the projected
mass of the cluster within the same radius that is not associated with
the main centrally-located dark matter halo and cluster central
galaxy. A complementary analysis of archival \chandra and \emph{ASCA}
observations of eight and one of these clusters respectively also
provided accurate measurements of the temperature of each cluster
($\txtot$) within a projected radius of $r\le1\hMpc$.  We refer the
reader to S03 for further details of the modeling and analysis of
these clusters.

We plot S03's mass and temperature measurements in Fig.~1.  The open
symbols show the individual clusters, and the filled symbols indicate
the properties of the mean relaxed and unrelaxed cluster sub-samples.
The mean temperatures of the relaxed and unrelaxed clusters, are
$\langle\txtot\rangle=6.3\pm0.8\keV$ and
$\langle\txtot\rangle=9.2\pm1.2\keV$ respectively, where the error
bars are the uncertainties on the means and are the estimated from
bootstrap re-sampling each sub-sample of clusters; we therefore expect
these error bars to be over-estimates.  The unrelaxed clusters appear
to be systematically hotter than the relaxed clusters.

Two of S03's sample (A\,383 -- Smith et al.\ 2001; A\,1835 -- e.g.\
Schmidt et al.\ 2001) have central cooling timescales of
$\tcool\ls10^9{\rm years}$.  This is in line with expectations from
other representative samples of X-ray luminous clusters (Peres et al.\
1998). S03 therefore recalculated all of the cluster temperatures
using an $0.05\le r\le1\hMpc$ annulus (i.e. excluding the cold core of
the two extreme ``cooling flow'' systems).  They found that, while the
temperature difference is slightly reduced ($\langle\txann\rangle_{\rm
relaxed}=6.9\pm0.9\keV$), it is robust to the exclusion of the central
$50\hkpc$ of each cluster from the temperature calculations.  The
$30\%$ temperature difference therefore does reflect a bona fide
difference between the ambient temperatures of relaxed and unrelaxed
clusters.  We note that this affect is similar to a
substructure-related bias found by Ota \& Mitsuda (2002) in the
cluster mass-luminosity relation.

\centerline{\psfig{file=figs/masstx.ps,width=75mm,angle=-90}}
\smallskip
\noindent{\small\addtolength{\baselineskip}{-5pt}
{\sc Fig.~1.} ---
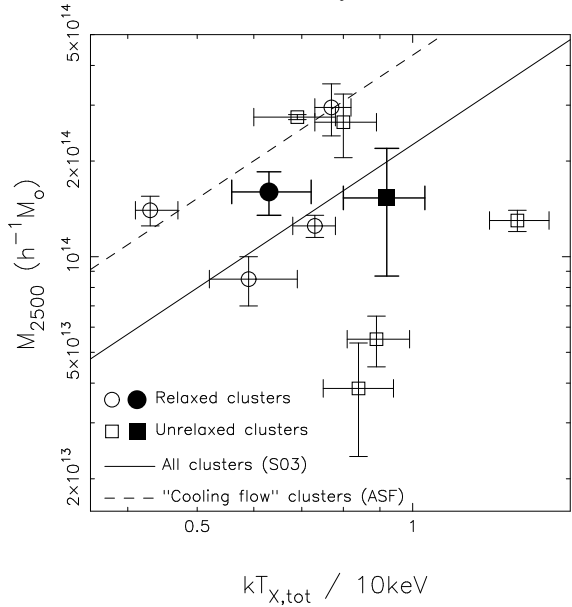
$M_{2500}$, the projected mass within $r_{2500}$, versus the
temperature of the intra-cluster medium for S03's sample of X-ray
luminous clusters.  The unrelaxed clusters are on average $30\%$
hotter than the relaxed clusters, causing a previously unquantified
structural bias in the normalization of the cluster mass-temperature
relation.  We also plot S03's and ASF's 
mass-temperature relations, assuming a canonical logarithmic slope of
$\alpha=2/3$ for the S03 relation.  The ASF relation agrees with the
two cooling flow clusters in S03's sample 
(A\,383 and A\,1835 are the two open circles that lie within
1--$\sigma$ of the line).

}
\setcounter{figure}{1}

\section{Modeling and Results}

\subsection{Approach}

We construct a simple model to investigate the impact of S03's results
on estimates of $\sig8$.  We begin by parameterizing the cluster
mass-temperature relation: 
\begin{equation}
\txtot(\keV)=A(M_{2500}/10^{14}\,h^{-1}\!\Msol)^\alpha
\end{equation}
where $\txtot$ and $M_{2500}$ are as defined in \S2, and $A,\alpha$
are the normalization and logarithmic slope respectively.  We first
convert the Jenkins et al.\ (2001) mass function to a temperature
function.  This conversion includes the following elements: a
mass-dependent concentration index (Eke, Navarro \& Steinmetz 2001),
conversion of three-dimensional masses from the simulations to
projected two-dimensional masses (Hjorth et al.\ 1998) and an
observational mass-temperature normalization (S03, ASF).  We then
fit this model temperature function to the observed temperature
function (Edge et al.\ 1990) using a single free parameter, $\sig8$.

Our model also contains the following parameters: \{$\om0$, $\Lo$,
$\Gamma$, $\sigma_T$\} where $\om0$ and $\Lo$ are the matter and
vacuum energy densities of the Universe at $z=0$, $\Gamma$ is the
spectral shape parameter for the power spectrum and $\sigma_T$ is the
scatter in $\log(\txtot$).  We focus our attention on the dependence
of $\sig8$ on $A$ and to a lesser extent on $\sigma_T$; we therefore
adopt ``concordance'' values for the remaining parameters:
$\om0=0.3$, $\Lo=0.7$, $\Gamma=0.2$ (e.g.\ Efstathiou et al.\ 2002),
$\alpha=2/3$ (e.g.\ ASF).  We stress that we adopt a fixed value of
$\om0=0.3$, and therefore do not investigate the $\sig8$--$\om0$
degeneracy.

\subsection{Model Fitting}

We use two independent mass-temperature calibrations to normalize our
models.  We begin with S03's normalization, and adopt the values of
$A$ and $\sigma_T$ relevant to their entire sample: $A=4.4$,
$\sigma_T=0.1$ (see solid line in Fig.~1).  We compute a model
temperature function and fit it to the observed temperature function
(Edge et al.\ 1990), obtaining a best-fit of $\sig8=0.75\pm0.05$ where
the uncertainty is the statistical uncertainty on the fit.  We plot
this best-fit model and the observed temperature function in Fig.~2.
Next, we turn to ASF's cooling flow mass-temperature relation.  These
authors observed a sample of seven cooling flow clusters with
\emph{Chandra}, and used these data to normalize the mass-temperature
relation.  We convert ASF's cooling flow mass-temperature relation
into the form required for our model: $A=2.6$, $\sigma_T\simeq0.03$.
Using these values, we construct a model temperature function and fit
it to the observed function, obtaining $\sig8=0.91\pm0.07$.  This
model (Fig.~2) fits the data less well than the S03-based model, with
the largest residuals occurring at high temperatures.

\smallskip
\centerline{\psfig{file=figs/tfunc.ps,width=85mm,angle=-90}}
\smallskip
\noindent{\small\addtolength{\baselineskip}{-5pt}
{\sc Fig.~2.} ---
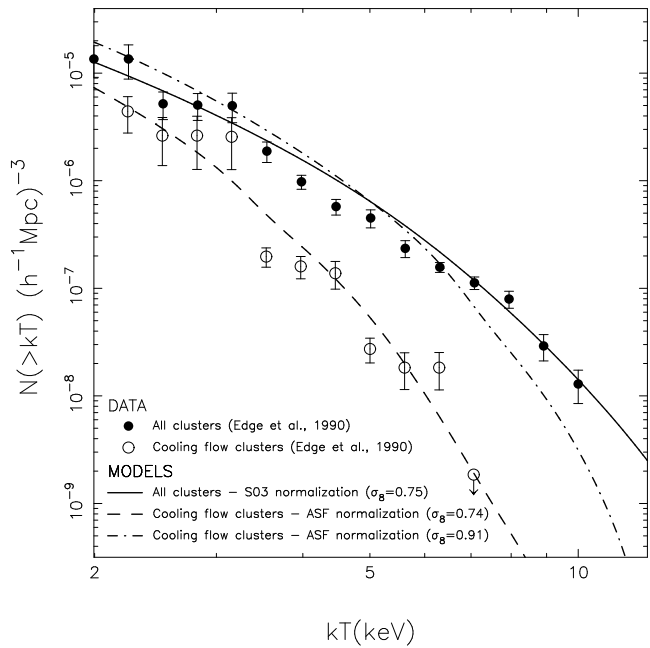
We plot the Edge et al.\ (1990) cluster temperature function for both
all clusters and cooling flow clusters (defined as containing a
line-emitting central galaxy), together with the best-fit model
temperature functions that are normalized with the S03 and ASF
mass-temperature relations.  When a cooling flow cluster based
normalization is applied to a representative sample of clusters,
$\sig8$ is over-estimated by $\sim20\%$.  However, when the cluster
selection function is accounted for properly in both the model
normalization and the observed temperature function, consistent values
of $\sig8$ are obtained.

}
\smallskip\setcounter{figure}{2}

A simple interpretation of these two models is that it is important to
understand the cluster selection function when using cluster
abundances to measure $\sig8$.  Specifically, using a cooling-flow
cluster mass-temperature normalization when studying the temperature
function of a representative sample of clusters may cause $\sig8$ to
be over-estimated by approximately $20\%$.

We test this interpretation by fitting the cooling flow normalized
model to a temperature function that describes just cooling flow
clusters.  We first use the observed correlation between line emission
from cluster central galaxies and short cooling timescales
($\tcool\ls10^9$ years -- Edge et al.\ 1992; Peres et al.\ 1998;
Crawford et al.\ 1999) to construct a ``cooling flow only''
temperature function from the Edge et al.\ (1990) sample.  We then fit
the cooling flow model to the cooling flow data and obtain a best-fit
value of $\sig8=0.74\pm0.05$, which agrees with the value obtained
from the original model that was normalized with S03's results.  We
plot this best-fit model and the relevant data in Fig.~2.  This model
confirms our interpretation that cluster substructure is an important
and previously unidentified systematic effect at the $20\%$ level when
using cluster abundances to constrain $\sig8$.

\subsection{Further Uncertainties}

The low value of $\sig8$ obtained in \S3.2 is similar to a number of
other recent results that favor $\sig8\sim0.6$--0.8 (e.g.\ Seljak
2001; Reiprich \& B\"ohringer 2001; Borgani et al.\ 2001; Allen et
al.\ 2002; Viana, Nichol \& Liddle 2002; Lahav et al.\ 2002; Brown et
al.\ 2002; Jarvis et al.\ 2002).  However, several other uncertainties
need to be investigated before a reliable conclusion on the value of
$\sig8$ from cluster abundance determinations can be drawn.

Firstly, we highlight the extrapolation of S03's lens models from
$r_{2500}$ (i.e.\ approximately the edge of the \emph{HST}/WFPC2 field
of view at $z\sim0.2$) to the cluster virial radii as a key systematic
uncertainty in our analysis.  Bardeau et al.\ (in prep.)  investigate
this effect in detail through their weak-shear analysis of panoramic
($28'\times 42'$) CFH12k $BRI$--band imaging of the S03 cluster
sample.  Prior to the completion of this wide-field analysis, we note
that weak lensing analyses of individual clusters (e.g.\ King, Clowe
\& Schneider 2002) are unable to discriminate between isothermal
($\rho\propto r^{-2}$) and Navarro, Frenk \& White (1997)
($\rho\propto r^{-3}$) profiles on large scales.  We therefore exploit
this lack of discriminatory power to make a conservative estimate of
this systematic uncertainty.  We integrate both profiles over the
radial range $0.25\ls r\ls1.5\hMpc$ (i.e.\ the dynamic range over
which we are extrapolating), and estimate that the uncertainty in
profile shape introduces an uncertainty in the virial mass estimate
for an individual cluster of $\sim30\%$, which translates into an
uncertainty in cluster temperature (assuming $M\!\propto\,\tx^{3/2}$)
of $\sim20\%$.  This equates to a $\sim10\%$ ``extrapolation''
systematic uncertainty in $\sig8$.

Secondly, we identify the projection of three-dimensional cluster
masses from numerical simulations to observed two-dimensional masses
(\S3.1) as a further source of systematic uncertainty.  As Hjorth et
al.\ (1998) discuss, the magnitude of this uncertainty depends on the
slope of the cluster density profile at small radii.  Recent
observational results (Smith et al.\ 2001; Sand, Treu \& Ellis 2002;
Dahle, Hannestad \& Sommer-Larsen 2002) indicate that there may be
substantial intrinsic scatter in this slope, appearing to contradict
theoretical claims for a universal profile (e.g.\ Navarro, Frenk \&
White 1997).  Given these complications, we conservatively adopt a
further $10\%$ ``projection'' systematic uncertainty in $\sig8$.

In summary, although S03's detailed lens models allow the
``substructure'' systematic to be accounted for properly and (to first
order) eliminated from our analysis, ``extrapolation'' and
``projection'' uncertainties combine to produce a $\sim20\%$
systematic that we are unable to control with the current dataset.

\section{Summary and Discussion}

We have used S03's substructure-dependent cluster mass-temperature
normalization to investigate the impact of cluster substructure on
estimates of $\sig8$.  We find that when a cooling flow cluster
mass-temperature normalization is applied to the general cluster
population, $\sig8$ is over-estimated by $20\%$.  A clear
understanding of the cluster selection function is therefore
fundamental to attempts to constrain $\sig8$ with cluster abundances.
The simple X-ray luminosity-limited selection of S03's sample (\S2)
enable us to account for this ``substructure'' systematic from our
analysis and thus to estimate that $\sig8=0.75\pm0.05({\rm
statistical})$.  However, before we conclude that $\sig8=0.75$, we
highlight two further systematic effects which may bias our analysis:
extrapolation of S03's small field-of-view lens models out to the
cluster virial radii, and uncertainties in the relationship between
three-dimensional mass information contained in numerical simulations
and the two-dimensional mass information that is available from
observations.  We estimate conservatively that these effects combine
to produce a further $20\%$ systematic uncertainty, and therefore we
conclude from the present data that $\sig8=0.75\pm0.05({\rm
statistical})\pm0.15({\rm systematic})$.  We also note that the
recently reported discrepancies between \emph{XMM}- and
\emph{Chandra}-based cluster temperature measurements (Schmidt et al.\
2001; Majerowicz et al.\ 2002; Markevitch 2002) may introduce further
uncertainty into cluster abundance determinations of $\sig8$.

Our 20\% ``substructure'' systematic is similar to the discrepancy
between the canonical value of $\sig8\sim0.9$--1 (e.g.\ Eke, Cole \&
Frenk 1996; Pierpaoli et al.\ 2001; Bacon et al.\ 2002; Bond et al.\
2002; Hoekstra et al.\ 2002; Refregier et al.\ 2002; van Waerbeke et
al.\ 2002) and recent claims for $\sig8\sim0.6$--0.8 (Seljak 2001;
Reiprich \& B\"ohringer 2001; Borgani et al.\ 2001; Allen et al.\
2002; Brown et al.\ 2002; Jarvis et al.\ 2002; Lahav et al.\ 2002;
Schuecker et al.\ 2002; Viana, Nichol \& Liddle 2002).  Our results
therefore offer a physically motivated explanation for some of this
discrepancy.  Independent confirmation of this comes from Randall et
al.'s (2002) semi-analytic study of the effect of cluster mergers on
the observed luminosity and temperature functions, and thus on the
inferred cluster mass function.  Randall et al.\ predict that cluster
mergers boost the observed temperature function and can cause $\sig8$
to be over-estimated by $20\%$ if hydrostatic equilibrium is assumed
for non-equilibrium clusters, in agreement with our observational
results.

All three systematics discussed in this letter (substructure,
extrapolation and projection) affect the ability of cluster abundance
techniques to measure $\sig8$ accurately, regardless of whether
gravitational lensing or X-ray techniques are used to measure the
cluster masses.  However, the insensitivity of gravitational lensing
to the physical nature and state of the cluster matter means that a
combined strong- and weak-lensing study of a large, objectively
selected sample of clusters should be the tool of choice for future
cluster abundance studies.

\section*{Acknowledgments}

We thank Steve Allen, Andrew Benson, Harald Ebeling, Richard Ellis,
Andy Fabian, Carlos Frenk, Andrew Liddle, Pasquale Mazzotta, Tommaso Treu 
and Pedro Viana for useful discussions and assistance.  GPS 
acknowledges a postgraduate studentship from PPARC.  ACE, VRE and IRS
acknowledge University Research Fellowships from the Royal Society.
RCN thanks the Department of Physics at the University
of Durham for their hospitality during the summer of 2002 when this
work was performed, and is grateful for financial support
under grant number NAG-5606. IRS acknowledges a Philip Leverhulme
Prize Fellowship.  JPK acknowledges financial support from CNRS.

\end{document}